\documentclass[prb,a4paper,twocolumn,floatfix,showpacs,showkeys,amsmath,amssymb,nobibnotes,altaffilletter,superscriptaddress]{revtex4-2}

\usepackage{graphicx}
\usepackage{pslatex}
\usepackage{xspace}
\usepackage{subfigure}
\usepackage[utf8]{inputenc}
\usepackage[T1]{fontenc}
\usepackage{amsmath}
\usepackage{textcomp}
\usepackage{color}

\begin{document}

\title{Open-Circuit Voltage Limitation by Surface Recombination in Perovskite Solar Cells}

\author{Sebastian Reichert}
\affiliation{Institut für Physik, Technische Universität Chemnitz, 09126 Chemnitz, Germany}
\author{Katelyn Goetz}
\affiliation{Integrated Centre for Applied Physics and Photonic Materials and Centre for Advancing Electronics Dresden (cfaed), Technical University of Dresden, Nöthnitzer Straße 61, 01187 Dresden, Germany}
\author{Christopher Wöpke}
\affiliation{Institut für Physik, Technische Universität Chemnitz, 09126 Chemnitz, Germany}
\author{Yana Vaynzof}
\affiliation{Integrated Centre for Applied Physics and Photonic Materials and Centre for Advancing Electronics Dresden (cfaed), Technical University of Dresden, Nöthnitzer Straße 61, 01187 Dresden, Germany}
\author{Carsten Deibel}
\thanks{Corresponding author: Carsten Deibel, deibel@physik.tu-chemnitz.de}
\affiliation{Institut für Physik, Technische Universität Chemnitz, 09126 Chemnitz, Germany}

\date{\today}

\begin{abstract} 

Fundamental electronic processes such as charge-carrier transport and recombination play a critical role in determining the efficiency of hybrid perovskite solar cells. The presence of mobile ions complicates the development of a clear understanding of these processes as the ions may introduce exceptional phenomena such as hysteresis or giant dielectric constants. As a result, the electronic landscape, including its interaction with mobile ions, is difficult to access both experimentally and analytically. To address this challenge, we applied a series of small perturbation techniques including impedance spectroscopy (IS), intensity-modulated photocurrent spectroscopy (IMPS) and intensity-modulated photovoltage spectroscopy (IMVS) to planar $\mathrm{MAPbI_3}$ perovskite solar cells. Our measurements indicate that both electronic as well as ionic responses can be observed in all three methods and assigned by literature comparison. The results reveal that the dominant charge-carrier loss mechanism is surface recombination by limitation of the quasi-Fermi level splitting. The interaction between mobile ions and the electronic charge carriers leads to a shift of the apparent diode ideality factor from 0.74 to 1.64 for increasing illumination intensity, despite the recombination mechanism remaining unchanged.

\end{abstract}

\pacs{}

\keywords{}

\maketitle


\section{Introduction}

The numerous examples of application of hybrid perovskites in solar cells,\cite{Park2019} light-emitting diodes,\cite{Xu2019} field-effect transistors,\cite{Wu2019} lasers\cite{Stylianakis2019} and other (opto)electronic devices demonstrate the vast potential of these materials due to their advantageous electrical and optical properties.\cite{Fu2020,Chouhan2020} In the development of solar cells, hybrid perovskites are of particular interest as an inexpensive material\cite{Song2017} with remarkable absorption properties\cite{Fujiwara2018} and performance.\cite{AbdMutalib2018} Hybrid perovskites can be used as a highly efficient single junction photovoltaic technology or in combination with silicon as a tandem solar cell.\cite{Li2020,AlAshouri2020} Several studies suggest that hybrid perovskites are mixed ionic--electronic semiconductors showing phenomena such as hysteresis or giant dielectric effects.\cite{Liu2019,JuarezPerez2014} Although the efficiency of perovskite solar cells has advanced remarkably,\cite{Nrel} fundamental questions concerning the interplay of ions and electronic charge carriers are still under debate.\cite{Mosconi2016} In addition to hysteresis, mobile ions are held responsible for band bending, the accumulation of charge carries near interfaces and modifying charge-carrier injection.\cite{LopezVaro2018,Gottesman2016,Ebadi2019,Li2020_2} Interestingly, basic device properties such as their open-circuit voltage were found to be influenced by the properties of ionic defects, which in turn depend on the details of perovskite processing and device architecture.\cite{Reichert2020,reichert2020probing} Consequently, the mutual impact of ions on electrical charge carriers and vice versa limits the elucidation of the electronic properties of perovskite solar cells. For example, charge-carrier transport in perovskite solar cells was investigated by Herz and co-workers.\cite{Herz2017} The charge-carrier mobility was found to strongly depend on the device architecture, the perovskite material and its exact stoichiometric composition. Recombination dynamics of charge carriers in hybrid perovskites have often been investigated using photoluminescence spectroscopy, with the dominant recombination mechanisms depending on the photo-excitation densities.\cite{Stranks2014,Johnston2015} Interestingly, the low phonon energy in hybrid perovskites was attributed to be a critical factor for nonradiative recombination.\cite{Kirchartz2018} However, it is well-known that photoluminescence of perovskite materials is influenced by various factors -- including the measurement conditions and environment -- thus complicating the extraction of fundamental information regarding the recombination processes in perovskite materials.\cite{Goetz2020}

\begin{figure}[t]
  \includegraphics*[scale=0.8]{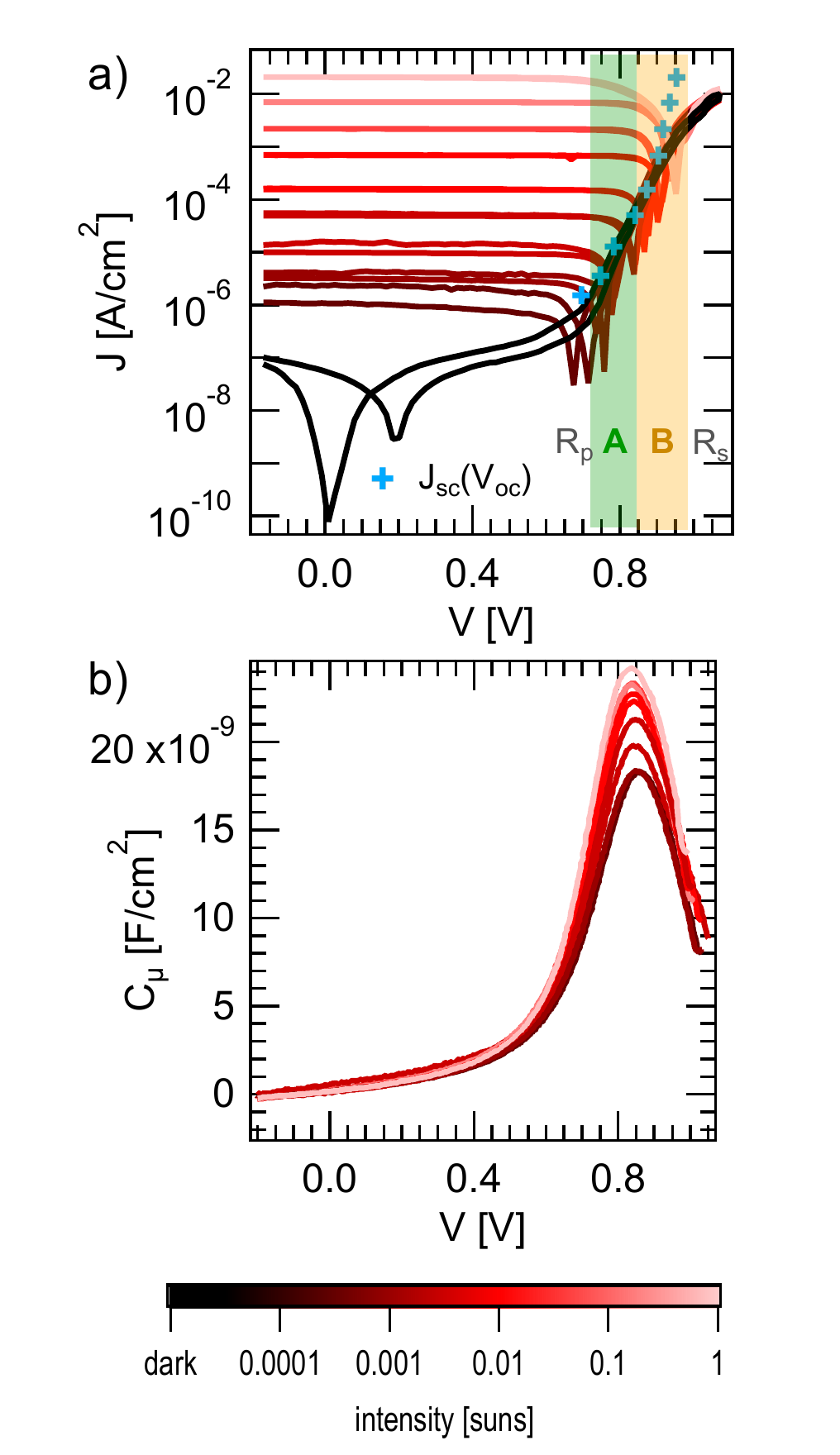}
  \caption{a) Current--density voltage characteristics with $J_\mathrm{sc}(V_\mathrm{oc})$ pairs and b) chemical capacitance versus voltage for different light intensities ranging from 0.0001~sun to 1~sun. Different regimes have been colour-coded as follows: $V<0.71~\mathrm{V}$: dominant $R_\mathrm{p}$; $0.71~\mathrm{V}<V<0.85~\mathrm{V}$: ionic influence on quasi Fermi level splitting (A); $0.85~\mathrm{V}<V<0.98~\mathrm{V}$: surface recombination (B); $V>0.98~\mathrm{V}$: dominant $R_\mathrm{s}$.}
  \label{fig:results_1}
\end{figure}
Small-perturbation techniques such as impedance spectroscopy (IS), intensity-modulated photocurrent spectroscopy (IMPS) and intensity-modulated photovoltage spectroscopy (IMVS) are powerful methods for the investigation of charge-carrier dynamics in semiconducting materials. In contrast to IS, in which the sample is excited by an AC-voltage modulation, a light-intensity modulation $P=P_0+P_\mathrm{ac}\mathrm{sin}(\omega t)$ is employed for IMPS and IMVS. The photocurrent $\Delta I$ of the solar cell is tracked during IS and IMPS measurements, while for IMVS, the photovoltage $\Delta V$ is recorded. For different DC intensities, the frequency $\omega$ of the AC component is also varied. The measured photocurrent and photovoltage have the same frequency, but are usually phase ($\phi$) shifted with a different amplitude. The transfer functions for IMPS and IMVS are as follows:
\begin{align}
    \label{eq:TF_IMPS_IMVS}
    Z(IMPS)= \frac{\Delta I}{P}\mathrm{e}^{\mathrm{i}\phi}, \qquad& 
    Z(IMVS)= \frac{\Delta V}{P}\mathrm{e}^{\mathrm{i}\phi}.
\end{align}
With IS, the frequency dependence of the capacitance $C(\omega)$ can be calculated from the impedance $Z=\Delta V/ \Delta I$ by selecting a suitable equivalence model, the simplest one being:
\begin{equation}
    \label{eq:C_Z}
    C=\frac{\mathrm{Im}\left(\frac{1}{\underline{Z}}\right)}{\omega}.
\end{equation}
IMPS and IMVS methods were previously employed to the study of recombination and transport processes of charge carriers in various types of solar cells.\cite{Ravishankar2019,Basham2014,Halme2011,Heiber2018} In addition to their time domain counterparts, such as transient photovoltage/photocurrent decay (TPV/TPC), IMPS and IMVS gain in significance for the investigation of hybrid perovskite solar cells. For example, Pocket et al.\ investigated a series of planar perovskite solar cells with IMPS, IMVS and TPV and showed that with all methods consistent values for the diode ideality factors can be obtained.\cite{Pockett2015} Another important work on the understanding and interpretation of intensity-modulated spectroscopy on perovskite solar cells is the study by Bernhardsgrütter et al.\cite{Bernhardsgrtter2019} The authors employed simulations to show that in addition to electronic transport and recombination, the transport of ions is visible in the spectra at low modulation frequencies. Additionally, various simulation parameters such as illumination intensity, charge-carrier mobilities, Shockley-Read-Hall lifetimes, ion densities and surface recombination velocities were varied and their influence on the resulting IMPS and IMVS spectra was examined. Ravishankar et al.\ developed an equivalent circuit model to find a basic explanation for the response of intensity modulated spectroscopic measurement for different perovskite materials and contact layers as well as for the interaction of electronic with ionic charge carriers.\cite{Ravishankar2019,Ravishankar2019_2} Importantly, the study by Chen et al.\ on planar perovskite solar cells indicates that IMVS provides information about a thermally activated recombination process of free charge carriers with homogeneously accumulated ones at the hole-transport layer interface.\cite{Chen2018} This result is confirmed by the study of Guill\'{e}n et al.\ with IMVS on perovskite solar cell fabricated in a mesoporous n-i-p architecture.\cite{Guilln2014}

In this study, the previous indications that surface recombination is dominant in perovskite solar cells is confirmed and validated by measuring current density--voltage characteristics (J--V), IS, IMPS and IMVS on planar MAPbI\textsubscript{3} perovskite solar cells (detailed processing conditions can be found in Sec.~\ref{sec:methods}). We utilise the results of these measurements to significantly expand this explanation by providing a comprehensive discussion of the solar cell ideality factor. Our results indicate that injection barriers -- presumably caused by the accumulation of ions at the interfaces --  limit the open-circuit voltage ($V_\mathrm{oc}$) of the devices. This interpretation is supported by a comparison of IS, IMPS and IMVS time constants with ionic migration rates from our previous work\cite{Reichert2020,reichert2020probing} and those reported in literature.


\section{Results and Discussion}\label{sec:results}

\begin{figure}[t]
  \includegraphics*[scale=0.75]{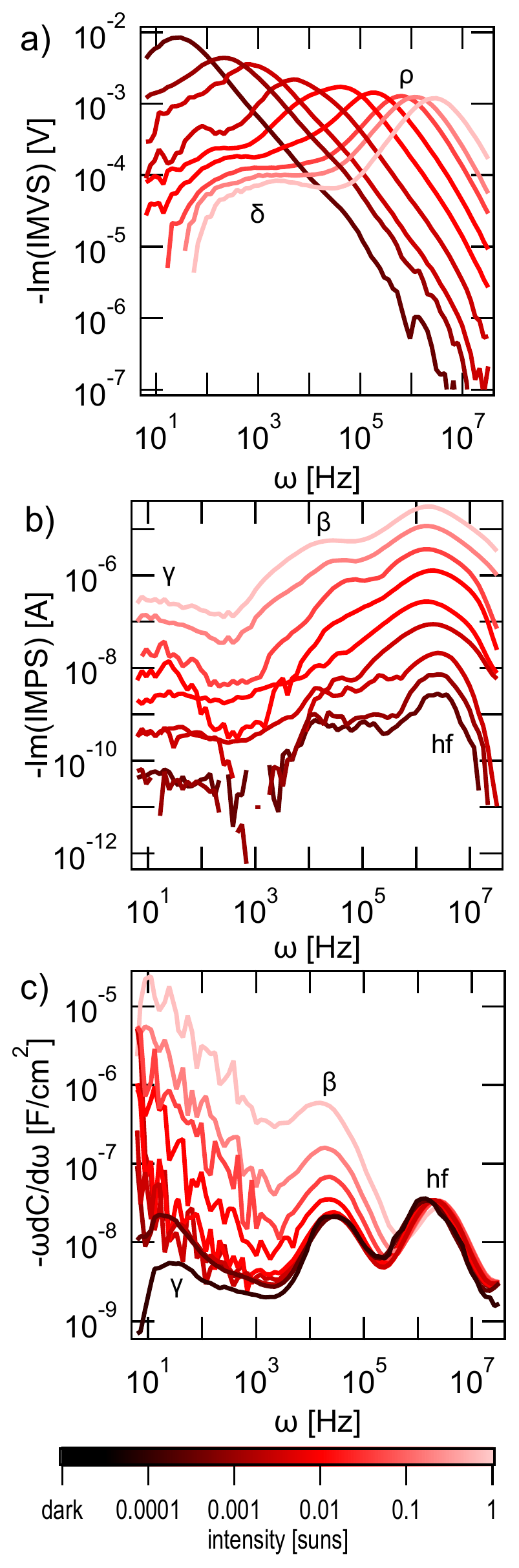}
  \caption{a) Intensity-modulated photovoltage spectroscopy (IMVS), b) intensity-modulated photocurrent spectroscopy (IMPS), c) derivation $-\omega\mathrm{d}C/\mathrm{d}\omega$ of IS measurements (Fig.~\ref{fig:Cf}) for different light intensities ranging from 0.0001 sun to 1 sun.}
  \label{fig:results_2}
\end{figure}

To acquire a complete data set via our multi-technique approach, a series of measurements of J--V (Fig.~\ref{fig:results_1}a) characteristics, capacitance--voltage measurements (CV) (Fig.~\ref{fig:results_1}b and \ref{fig:CV}), IMVS (Fig.~\ref{fig:results_2}a), IMPS (Fig.~\ref{fig:results_2}b) and IS (Fig.~\ref{fig:Cf}) with light intensities varying over five orders of magnitude were performed. CV measurements were done by a fast sweeping method, where a pre-bias of 1.2 V is applied to the solar
cell for one minute.\cite{Fischer2018} Detailed measurement conditions can be found in Sec.~\ref{sec:methods}. The $J_\mathrm{sc}(V_\mathrm{oc})$ values extracted from the J--V curves measured under illumination fit well with the measurements performed in the dark, which motivates the calculation of the ideality factor as will be discussed later. To obtain the chemical capacitance $C_\mu$ (Fig.~\ref{fig:results_1}b), the geometrical capacitance $C_\mathrm{geo}$ has to be subtracted from the CV measurements (Fig.~\ref{fig:CV}). A small shift in the peak position and height of the injection capacitance can be observed, which is linked to the change in the charge-carrier density due to illumination. The chemical capacitance enables the calculation of the charge-carrier density within the active layer, as described below. In the case of IMVS, two peaks can be observed, which were labelled with $\rho$ and $\delta$. The illumination dependence of the $\rho$ peak extends over the entire frequency range, whereas $\delta$ shows only a small dependence on light intensity. An approximately linear increase in amplitude for both peaks was measured with increasing illumination. Three peaks are visible for IMPS, which were labelled with $hf$, $\beta$ and $\delta$. The high-frequency $hf$ response shows a negligible illumination dependence in contrast to the $\beta$ and $\gamma$ peaks. For evaluating the IS data, the derivation $-\omega\mathrm{d}C/\mathrm{d}\omega$ was calculated as shown in Fig.~\ref{fig:results_2}c. Peaks in Fig.~\ref{fig:results_2}c are correlated to ionic migration rates,
\begin{equation}
    \label{eq:e_t}
    e_\mathrm{t}=\omega_\mathrm{max}.
\end{equation}
Because of the agreement in the peak positions of IS with those observed via IMPS, these three peaks were also labelled with $hf$, $\beta$ and $\gamma$. Notably, the responses $\beta$ and $\gamma$ show a strong dependence on illumination in both position and amplitude, in contrast to the $hf$ peak.


\begin{figure}[t]
  \includegraphics*[width=\columnwidth]{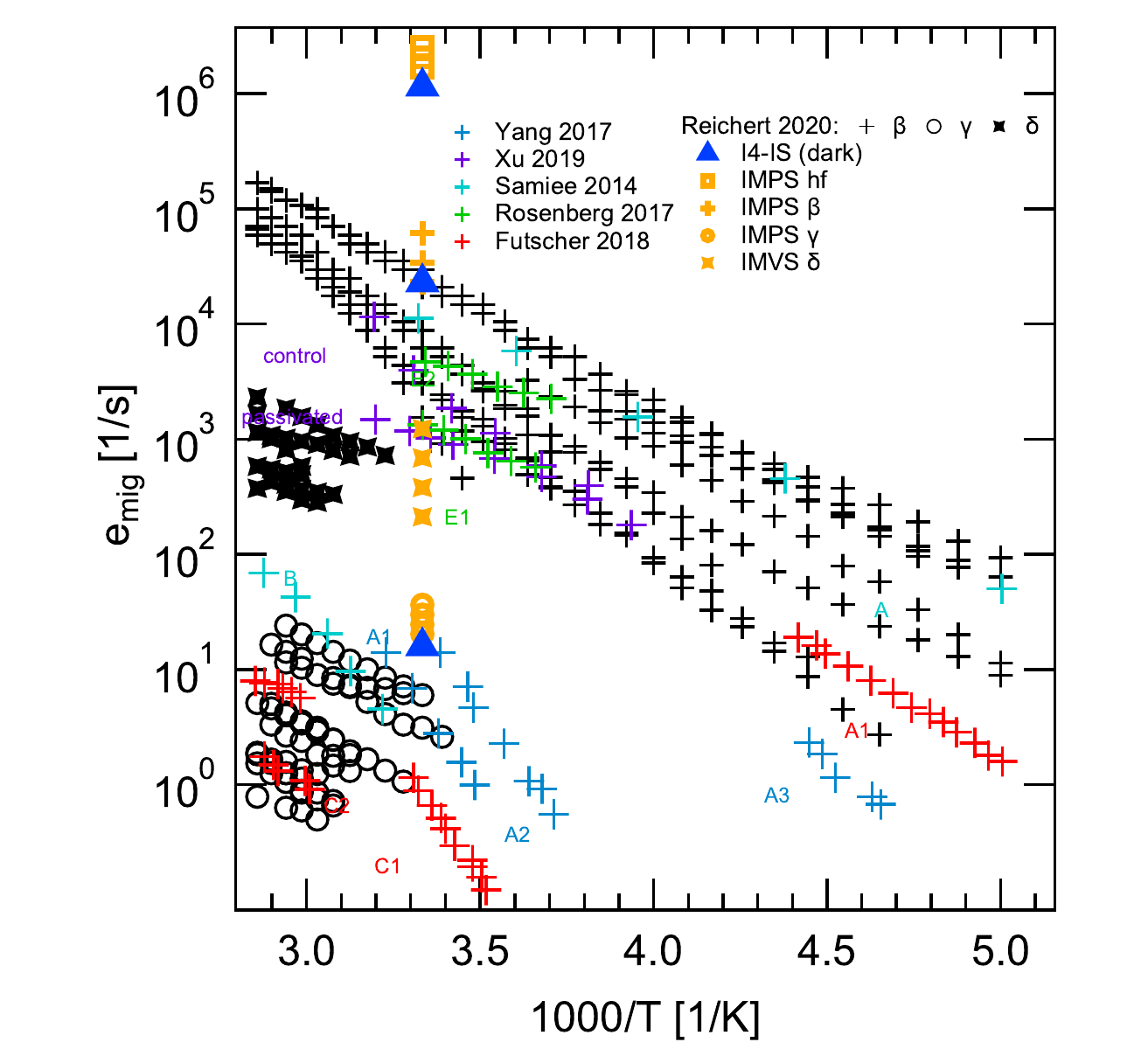}
  \caption{Comparison of ionic migration rates determined by IS, IMPS and IMVS with values from literature and our recent work (Reichert 2020\cite{reichert2020probing}).}
  \label{fig:Masterarrhenius}
\end{figure}

To identify the peaks observed in the IS, IMPS and IMVS measurements (Figs.~\ref{fig:results_2}a, \ref{fig:results_2}b and \ref{fig:results_2}c), we compared the ionic migration rates calculated from their peak positions with those from our previous work\cite{Reichert2020,reichert2020probing} and other reports found in literature using an Arrhenius diagram (Fig.~\ref{fig:Masterarrhenius}).  As shown in Fig.~\ref{fig:Masterarrhenius}, the determined rates of $\beta$, $\gamma$ and $\delta$ from IS, IMPS and IMVS agree well with ionic migration rates from literature, including our results on a batch of identically fabricated solar cells with variation of the precursor stoichiometry.\cite{Reichert2020,reichert2020probing} We therefore conclude that the low-frequency responses $\beta$, $\gamma$ and $\delta$ can be attributed to the same ionic defects as observed in the previous studies. This result is in agreement with the simulations of Bernhardsgrütter et al.\cite{Bernhardsgrtter2019} and other studies.\cite{TurrenCruz2018,Domanski2017,Chen2018,Prochowicz2017,Roose2017} The movement of mobile ions during IMPS and IMVS measurements can be explained by the variation of the internal electrical field caused by changing the photoinduced free charge-carrier density. This change leads to a redistribution of mobile ions and is therefore in analogy to the explanation of ionic movement caused by the modulation of the Fermi level in IS. Additionally, the dependence of the peak height in IS (Fig.~\ref{fig:results_2}c), which is related to the ionic-defect density, supports this finding. The observable increase of the peak height in Fig.~\ref{fig:results_2}c can be explained by changing the thickness of the Debye layers formed by ion accumulation at the interfaces.\cite{reichert2020probing,Almora2015} The migration and redistribution itself can have an impact on the electronic charge carriers and vice versa, making the overall charge-carrier dynamics in perovskite solar cells rather complex.\cite{Bernhardsgrtter2019} Deviations between the migration rates obtained by IS in comparison to our previous works\cite{Reichert2020, reichert2020probing} (i.e.\ for $\gamma$) can likely be explained by small variations in the processing conditions. Differences between IS and IMPS/IMVS can be the result of different measurement methods, as they can probe different parts of the same ionic defect distribution.\cite{Reichert2020} 

It has been previously shown that the high-frequency response $hf$ in the IMPS spectra is dominated by the $RC$ time constant.\cite{Bernhardsgrtter2019, Pockett2015,Ravishankar2019_2,Prochowicz2017} This finding is supported by the negligible intensity dependence of the peak position in IMPS and IS. Consequently, no information concerning the charge-carrier transport in the perovskite layer can be obtained. We point out that the calculation of the charge-carrier mobility with the active layer thickness $L$ and the peak frequency $\omega_{hf}$ of the $hf$ response,\cite{Nojima2019}
\begin{equation}
    \label{eq:mobility}
    \mu=\frac{\omega_{hf}L^2}{2V_\mathrm{oc}},
\end{equation}
would result in a value of $10^{-3}~\mathrm{cm^2/(Vs)}$, which is at least two orders of magnitude lower than values reported in literature.\cite{Long2014,Herz2017} We therefore estimated the geometrical capacitance from the capacitive response at $\omega \approx 2\cdot 10^5~\mathrm{Hz}$ for further calculations, i.e.\ before the $hf$ response dominates the spectra. We have also chosen this frequency for the CV measurement in order to exclude the influence of mobile ions on the calculation of the charge-carrier density as discussed later. Since IMVS is measured at $V_\mathrm{oc}$, where no net current is flowing, the series resistance is not a critical parameter and the interpretation of the high-frequency response in IMVS remains valid. The IMVS $\rho$ response, with an intensity dependence over the entire frequency range, cannot be compared to the ionic migration rates from Fig.~\ref{fig:Masterarrhenius}. We therefore propose that the $\rho$ response is the result of recombination of charge carriers. To support this concept, in the following, a detailed analysis of the solar cells' ideality factor is provided, to determine the dominant recombination mechanism and the impact of mobile ions on the electronic landscape.

First, the calculation of the ideality factor from the measured $J_\mathrm{sc}(V_\mathrm{oc})$ values is considered. Starting from the well-established Shockley diode equation to describe the current--voltage characteristics of perovskite solar cells,\cite{Shockley1949}
\begin{equation}
    J(V)=J_0 \left( \mathrm{exp}\left(\frac{eV}{n_\mathrm{id}k_\mathrm{B}T}  \right)-1 \right)-J_\mathrm{sc},
    \label{eq:ideal_shockley_dark}
\end{equation}
where $J_0$ is the saturation current density, $e$ the elementary charge, and $k_\mathrm{B}T$ the thermal energy, an equation for determining the ideality factor by setting $J(V_\mathrm{oc})=0$ and $J_0 \ll J_\mathrm{sc}$ can be found:\cite{Wolf1963,Tvingstedt2016}
\begin{equation}
    \mathrm{ln}(J_\mathrm{sc})=\frac{eV_\mathrm{oc}}{n_\mathrm{id}k_\mathrm{B}T}+\mathrm{ln}(J_0).
    \label{eq:nid}
\end{equation}
Then $n_\mathrm{id}$ can be obtained by a fit of the $J_\mathrm{sc}(V_\mathrm{oc})$ values in Fig.~\ref{fig:results_1}a.
\begin{figure}[t]
  \includegraphics*[scale=0.8]{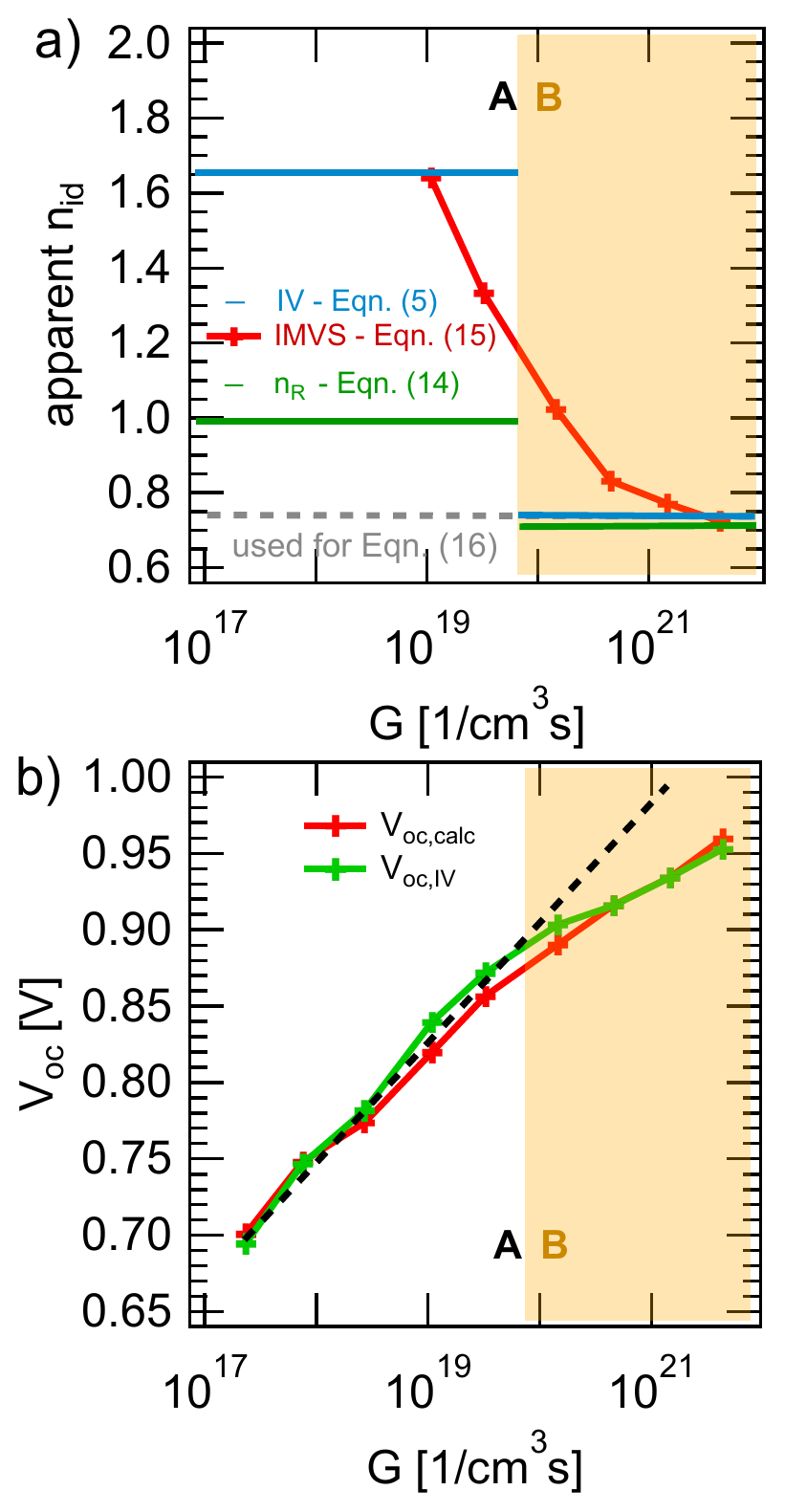}
  \caption{a) $n_\mathrm{id}$ calculated from real part of IMVS measurements according to Fig.~\ref{fig:Re_IMVS}, from $J_\mathrm{sc}(V_\mathrm{oc})$-values and $n_\mathrm{R}$ using Eqn.~\ref{eq:n_R}. J--V and partly $n_\mathrm{R}$ include the ionic response. Only the ideality factor found in regime B, applied also to regime A, allows a suitable $V_\mathrm{oc}$ reconstruction in b) across both regimes. b) Measured and reconstructed $V_\mathrm{oc}$ versus the generation rate $G$. The dashed line represent a linear curve as guide for the eyes.}
  \label{fig:discussion}
\end{figure}
A single fit over the entire voltage range results in a large fitting error. Therefore, two fits were made yielding values of 1.65 and 0.74 for the low and high illumination intensities (i.e.\ low generation rates $G$), respectively, as shown in Fig.~\ref{fig:discussion}a. We labelled the low-intensity regime with A and the high-intensity regime with B. Assuming a symmetric quasi-Fermi level splitting, the ideality factor is linked to the recombination order $\lambda+1$ as follows:\cite{Wolff2019}
\begin{equation}
    n_\mathrm{id}\approx \frac{2}{\lambda+1}
    \label{eq:n_id}
\end{equation}
leading to values between 2/3 (Auger recombination) and 1 (direct band-to-band recombination) to 2 (Shockley-Read-Hall recombination, SRH).\cite{Nelson2003,Wrfel2005} This would suggest that SRH recombination is dominant at low light intensities, while Auger is dominant at high intensities. However, Auger recombination was found to be unlikely for such low charge-carrier densities. Therefore, an alternative explanation for these low ideality factors -- surface recombination -- will be discussed later.

Next, the calculation of the ideality factor from intensity-modulated spectroscopy is considered. The charge-carrier density can be determined using the following equation:\cite{Heiber2018}
\begin{equation}
    n=\frac{1}{eL}\int_{V_\mathrm{sat}}^{V}C_\mu(V')\mathrm{d}V'+n_\mathrm{sat},
    \label{eq:n}
\end{equation}
where $V_\mathrm{sat}$ and $n_\mathrm{sat}$ refer to the saturation voltage and saturation charge-carrier density, respectively. $V_\mathrm{sat}$ is typically chosen to be in the reverse bias regime. $C_\mu$ is shown in Fig.~\ref{fig:results_1}b. $n_\mathrm{sat}$ can be estimated by:\cite{Proctor2013}
\begin{equation}
    n_\mathrm{sat}(V_\mathrm{sat})=\frac{1}{eL}C_\mu (V_\mathrm{sat})(V_\mathrm{oc}-V_\mathrm{sat}),
    \label{eq:n_sat}
\end{equation}
and
\begin{equation}
    V'=V-J(V)AR_\mathrm{s},
    \label{eq:V_sat}
\end{equation}
which accounts for the voltage drop caused by the series resistance. Here $A$ is the area of the active layer and $R_\mathrm{s}$ is the series resistance. The $n(V_\mathrm{oc})$ dependence yields a curve, which can by described by:\cite{Foertig2012}
\begin{equation}
    n(V_\mathrm{oc})=n_0 \mathrm{exp}\left(\frac{eV_\mathrm{oc}}{n_\mathrm{n}k_\mathrm{B}T} \right),
    \label{eq:n_Voc}
\end{equation}
where $n_0$ refers to the dark charge-carrier density at 0~V. Based on Eqn.~\ref{eq:n_Voc}, the ideality factor of the charge-carrier density is found to be $n_\mathrm{n}=6.0$ in regime A and $n_\mathrm{n}=7.8$ in regime B, as is shown in Fig.~\ref{fig:n_tau}a.

The charge-carrier lifetime $\tau$ can be obtained from the $\rho$ response peaks of the IMVS measurement:\cite{Guilln2014}
\begin{equation}
    \tau=\frac{1}{\omega_\mathrm{max}}.
    \label{eq:tau}
\end{equation}
The dependence of $\tau$ on $V_\mathrm{oc}$ follows:\cite{Foertig2012}
\begin{equation}
    \tau_\mathrm{n}(V_\mathrm{oc})=\tau_0 \mathrm{exp}\left(-\frac{eV_\mathrm{oc}}{n_\tau k_\mathrm{B}T} \right).
    \label{eq:tau_Voc}
\end{equation}
and fitting with Eqn.~(\ref{eq:tau_Voc}) yields the ideality factor of charge carrier lifetime in regime A to $n_\tau=1.18$ and for regime B to $n_\tau=0.82$. Here, $\tau_0$ is the dark charge carrier lifetime at 0~V. These two ideality factors can be combined to calculate the ideality factor of recombination $n_\mathrm{R}$:\cite{Foertig2012,Set2015}
\begin{equation}
    \frac{1}{n_\mathrm{R}}=\frac{1}{n_\mathrm{n}}+\frac{1}{n_\mathrm{\tau}}.
    \label{eq:n_R}
\end{equation}
For regime A, $n_\mathrm{R}$ equals 0.99, whereas for regime B $n_\mathrm{R}$ is 0.74. The latter value is consistent with the diode ideality factor calculated from $J_\mathrm{sc}(V_\mathrm{oc})$ values at high illumination intensities. For regime A, $n_\mathrm{R}$ is too low in comparison to the result from $J_\mathrm{sc}(V_\mathrm{oc})$ values. Calado et al.\cite{Calado2019} found that different voltage preconditions of the perovskite solar cell can lead to deviations in the apparent $n_\mathrm{id}$, i.e.\ values of 1 or 2 can be obtained although in both cases surface recombination is dominant. This effect was attributed to the redistribution of mobile ions. We note that our CV measurements were performed using a fast sweeping method,\cite{Fischer2018} where a prebias of 1.2~V for one minute is applied to the solar cell. The different preconditioning between CV and JV/IMVS might serve as explanation for the different apparent $n_\mathrm{id}$, if the impact of the redistribution of the mobile ions on the electronic landscape is taken into account, as will be discussed later. 

Another way to determine the ideality factor can be achieved by considering the real part of the IMVS measurement (Fig.~\ref{fig:Re_IMVS} and \ref{fig:discussion}a). Based on Eqn.~(\ref{eq:nid}), an equivalent equation using the change of $V_\mathrm{oc}$ upon modulation of the generation rate $G\approx J_\mathrm{sc}/eL$ can be found:
\begin{equation}
    n_\mathrm{id}\approx \frac{e}{k_\mathrm{B}T}\frac{\Delta V_\mathrm{oc}}{\Delta \mathrm{ln}(G)}.
    \label{eq:n_id_IMVS}
\end{equation}
Here, $\Delta V_\mathrm{oc}$ is given by $\mathrm{Re}(\mathrm{IMVS})(f_\mathrm{min})$ at the lowest measured frequency $f_\mathrm{min}$. This allows the calculation of $n_\mathrm{id}$ as shown in Fig.~\ref{fig:discussion}a using Fig.~\ref{fig:Re_IMVS}. Values of $G<10^{19}~\mathrm{1/cm^3s}$ were excluded since they would lead to unreasonably high $n_\mathrm{id}$ caused by the apparent recombination response $\rho$ at low frequencies in the IMVS spectra of Fig.~\ref{fig:Re_IMVS}. The values for $n_\mathrm{id}$ using Eqn.~(\ref{eq:n_id_IMVS}) indicate a shift from $n_\mathrm{id}=0.73$ for high illumination intensities to $n_\mathrm{id}=1.64$ for low illumination intensities and are in good agreement with the values from $J_\mathrm{sc}(V_\mathrm{oc})$ (see Fig.~\ref{fig:discussion}a).

According to the study of Tress et al.,\cite{Tress2018} values of $n_\mathrm{id}<1$ for high illumination intensity can be explained by considering the process of surface recombination. Charge carriers can be accumulated at the interfaces to the transport layers if potential barriers hinder their extraction. As a result, surface recombination can be the dominant mode of charge-carrier loss. This finding is confirmed by several studies,\cite{Guilln2014,Wheeler2015,Courtier2020} and especially by the study of Chen et al.\cite{Chen2018}, where a linear dependence between the recombination time constant and illumination intensity was found to support this conclusion. Experimental results and simulations by Sundqvist et al.\cite{Sundqvist2016} have shown that $n_\mathrm{id}$ is reduced to 2/3 if high injection barriers are present in the solar cell. An analytical explanation was provided. We therefore consider regime A in Figs.~\ref{fig:results_1}a and
~\ref{fig:discussion} to be dominated by surface recombination. This means that the quasi-Fermi level splitting is hindered by the work function of the contacts, which in turn reduces the $V_\mathrm{oc}$. In this case, $V_\mathrm{oc}$ should saturate for high illumination intensities.\cite{Tress2018} Such a saturation is shown in Fig.~\ref{fig:discussion}b. $V_\mathrm{oc}$ was taken from the JV measurement. Additionally, we reconstructed $V_\mathrm{oc}$ according to:\cite{Foertig2012}
\begin{equation}
    V_\mathrm{oc}=n_\mathrm{R}\frac{k_\mathrm{B}T}{e}\mathrm{ln}\left(\frac{R(n)}{R_0} \right),
    \label{eq:V_oc_IMS}
\end{equation}
with the recombination rates given by:
\begin{align}
    \label{eq:R_R_0}
    R=\frac{n(V)}{\tau_\mathrm{n}(V)},\qquad & 
    R_0=\frac{n_0}{\tau_\mathrm{n_0}},
\end{align}
and the recombination ideality factor $n_\mathrm{R}$ calculated from Eqn.~(\ref{eq:n_R}). $R_0$ is the dark recombination rate at 0~V. Both methods agree with each other and show an apparent saturation of $V_\mathrm{oc}$ indicating a limitation by injection barriers. According to the recent study of Li et al.,\cite{Li2020_2} an injection barrier can be introduced by mobile ions. It was shown that an accumulation of anions on the electron transport layer leads to a barrier, which hinders the electron extraction and injection.

We note that the agreement between measured and reconstructed $V_\mathrm{oc}$ is only possible when the constant value of $n_\mathrm{R}=0.74$, determined only for high illumination intensities, is used (regime B). We propose that surface recombination is the dominant recombination mechanism independent of the illumination intensity. The apparent $n_\mathrm{id}$ and $n_\mathrm{R}$ larger than 1 calculated by Eqn.~(\ref{eq:n_R}) and (\ref{eq:n_id_IMVS}), respectively, must therefore by the result of an ion related mechanism: Calado et al.\cite{Calado2019} found that a redistribution of mobile ions can lead to a change in the electron and hole population overlap at the interface regions. That means the redistribution of mobile ions can cause a change from symmetric to asymmetric quasi-Fermi level splitting, which results in a shift of the apparent ideality factor depending on the illumination intensity. We therefore attribute regime A in Fig.~\ref{fig:results_1}a and \ref{fig:discussion} to be the result of ion redistribution, i.e.\ interaction between mobile ions at the interfaces with the free charge carriers.

The limitation of $V_\mathrm{oc}$ by injection barriers should be observable in the low apparent built-in potentials calculated from CV measurements. The calculated built-in potentials according to the Mott-Schottky evaluation (Fig.~\ref{fig:CV}) by
\begin{equation}
    \label{eq:MS}
    1/C^2=\frac{2(V_\mathrm{bi}-V)}{e\epsilon_0\epsilon_\mathrm{R}N_\mathrm{eff}}
\end{equation}
show much higher values than $V_\mathrm{oc}$ and an apparent shift with illumination intensity. We assume that the built-in potential determined by the Mott--Schottky evaluation does not represent the real conditions within the solar cells, as the CV measurement is affected by mobile ions. Therefore, a correction of the built-in potential by a redistribution of the mobile ions with changing light intensity might by necessary, suppressing the shift of $V_\mathrm{bi}$, as discussed in our recent studies\cite{Reichert2020,reichert2020probing} and in detail by Ravishankar et al.\cite{Ravishankar2019,Ravishankar2019_2}.


\section{Summary}

We have experimentally shown that the slow, approximately light-independent responses in both IMVS and IMPS correspond to ion migration rates in correspondence with IS. Additionally, IMVS show a strong response over several orders of magnitude in light intensity, which can be related to charge-carrier recombination. We verified that surface recombination is the dominant loss mechanism and discussed the ideality factor of the solar cell in detail to support this finding. We showed that the apparent ideality factor can be influenced by the redistribution of mobile ions. This study provides a deeper understanding of the ionic and electronic charge-carrier dynamics in perovskite solar cells and how to improve their electrical parameters by identifying the dominant recombination mechanism of photogenerated charge carriers.

\section{Methods}\label{sec:methods}

\textbf{Processing:} The MAPbI\textsubscript{3} active layer was processed by the lead acetate trihydrate approach\cite{Zhang2015,An2019,Fassl2018} using a precursor solution with a stoichiometry of 1:3.00 MAI:PbAc\textsubscript{2} and an additive of hypophosphoric acid at a ratio of 1.7~$\mu$l/100 mg MAI. The perovskite solution was spin cast at 2000~rpm for 60~s in a dry-air filled glovebox (relative humidity $<0.5~\%$) on top of a modified poly(3,4-ethylene-dioxythiophene):poly(styrenesulfonate) (m-PEDOT:PSS) as hole-transport layer.\cite{Zuo2016} The hole-transport layer was spin cast on the cleaned substrates (Pre-patterned indium tin oxide (ITO) coated glass substrates (PsiOTech Ltd., $15~\Omega/\mathrm{sqr}$) ultrasonically cleaned with $2~\%$ Hellmanex detergent, deionized water, acetone, and isopropanol) at 4000~rpm for 30~s and annealed at $150~^\circ\mathrm{C}$ for 15~min.\cite{Zuo2016} The electron-transport layer ([6,6]-phenyl-C60-butyric acid methylester (PC\textsubscript{60}BM)) was dissolved in chlorobenzene (20~mg/ml) and dynamical spin cast at 2000~rpm for 30~s. After further annealing for 10~min. at $100~^\circ\mathrm{C}$, a  bathocuproine (BCP), 0.5~mg/ml dissolved in isopropanol, hole-blocking layer was spin cast on top of the PC\textsubscript{60}BM. Thermally evaporated silver was used as counter electrode, evaporated at a rate of 0.1-1 $\mathring{A}$/s to a thickness of 80~nm.

\textbf{Characterisation:} All measurements were performed using a Zurich Instruments MFLI lock-in amplifier with MF-IA, MF-MD and MF-5FM options and an Omicron A350 diode laser with fast analog modulation. The diode laser for AC and DC illumination was calibrated by measuring phase and amplitude over the entire frequency range using a photodetector (Newport 818-BB-21). The light intensity for 1~sun was determined using an AM 1.5 solar simulator with $100~\mathrm{mW/cm}^2$ irradiation (Wavelabs). To ensure a linear response during measuring IMPS and IMVS, the AC illumination-modulation amplitude was chosen to be $10~\%$ of the DC intensity. For achieving steady-state conditions, the samples were pre-illuminated for 30~s before each measurement. CV measurements were done applying an AC frequency of 80~kHz with an amplitude of $V_\mathrm{ac}=20~\mathrm{mV}$. For CV profiling, the solar cells were pre-biased at 1.2~V for 60~s and rapidly swept with 30~V/s in the reverse direction.\cite{Fischer2018}

\begin{acknowledgments}

Y.V.\ and C.D.\ thank the DFG for generous support within the framework of SPP 2196 project (PERFECT PVs). This project has received funding from the European Research Council (ERC) under the European Union's Horizon 2020 research and innovation programme (ERC Grant Agreement no.\ 714067, ENERGYMAPS).

\end{acknowledgments}


%


\widetext
\clearpage

\begin{center}
\textbf{\large Supplementary Information: Intensity Modulated Spectroscopy on Perovskite Solar Cells}
\end{center}
\setcounter{equation}{0}
\setcounter{figure}{0}
\setcounter{table}{9}
\setcounter{page}{1}
\setcounter{section}{0}
\makeatletter
\renewcommand{\theequation}{S\arabic{equation}}
\renewcommand{\thefigure}{S\arabic{figure}}
\renewcommand{\thetable}{S\arabic{table}}
\renewcommand{\bibnumfmt}[1]{[S#1]}
\renewcommand{\citenumfont}[1]{S#1}

\section{Capacitance--Voltage Measurements}

\begin{figure}[h]
  \includegraphics*[scale=0.8]{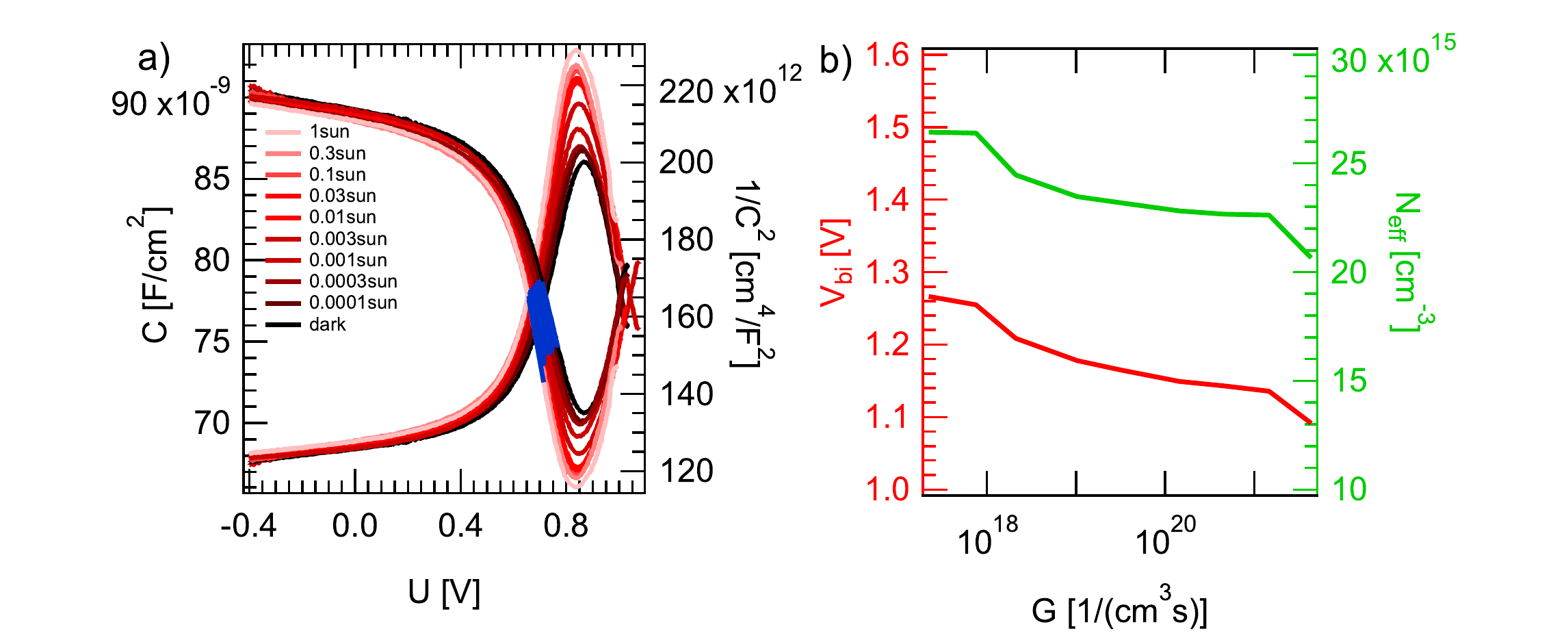}
  \caption{a) Capacitance--Voltage (CV) measurements and Mott--Schottky evaluation for all intensities. Blue curves in the depletion zone represent linear fits for determining built-in potential $V_\mathrm{bi}$ and effective doping density $N_\mathrm{eff}$. b) $V_\mathrm{bi}$ and $N_\mathrm{eff}$ versus generation rate $G$.}
  \label{fig:CV}
\end{figure}

\section{Capacitance--frequency measurements (Cf)}

\begin{figure}[h]
  \includegraphics*[scale=0.6]{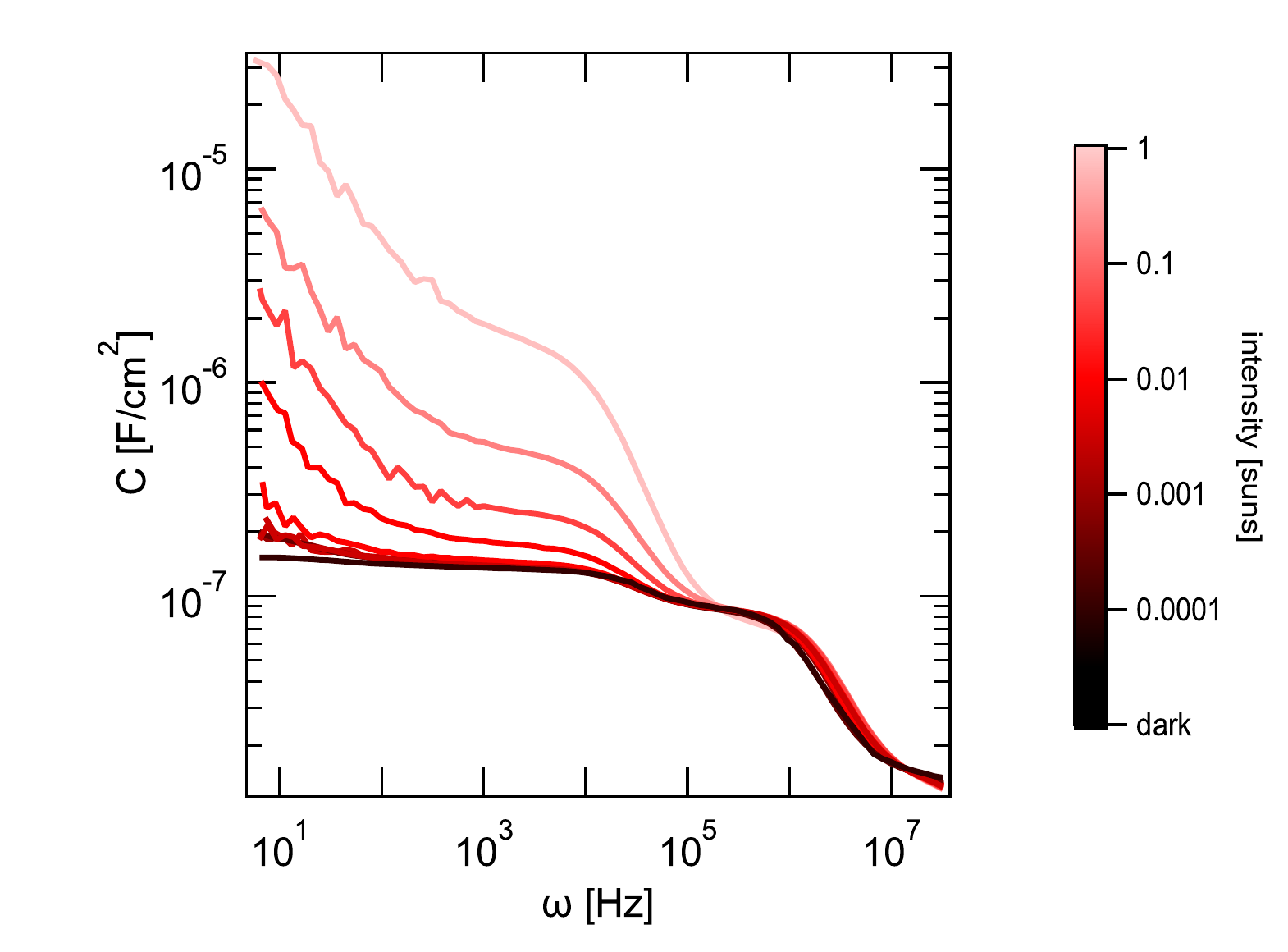}
  \caption{Capacitance--frequency measurements (Cf) for different illumination intensities.}
  \label{fig:Cf}
\end{figure}

\newpage

\section{Charge Carrier Density and Lifetime}

\begin{figure}[h]
  \includegraphics*[scale=0.8]{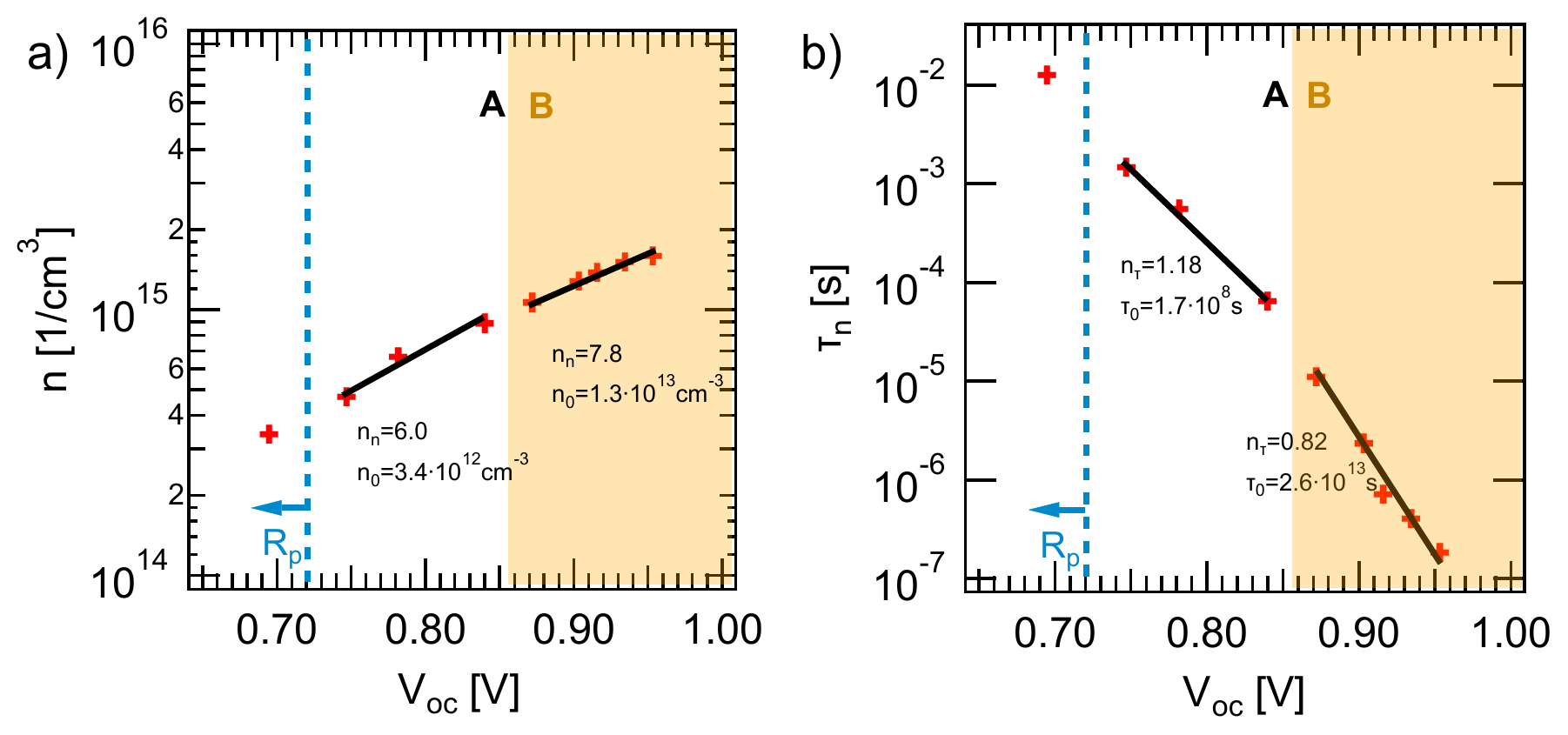}
  \caption{a) Charge carrier density $n$ extracted from CV measurements and intensity-modulated photocurrent spectroscopy (IMPS). b) Charge carrier lifetime calculated from intensity-modulated photovoltage spectroscopy (IMVS). The ideality factors $n_\mathrm{n}$ and $n_\mathrm{\tau}$ can be extracted from the slope of linear fits whereby the charge carrier density $n_0$ and lifetime $\tau_0$ in dark can be obtained by the intersection point with the ordinate.}
  \label{fig:n_tau}
\end{figure}

\section{Real Part of IMVS}

\begin{figure}[h]
  \includegraphics*[scale=0.6]{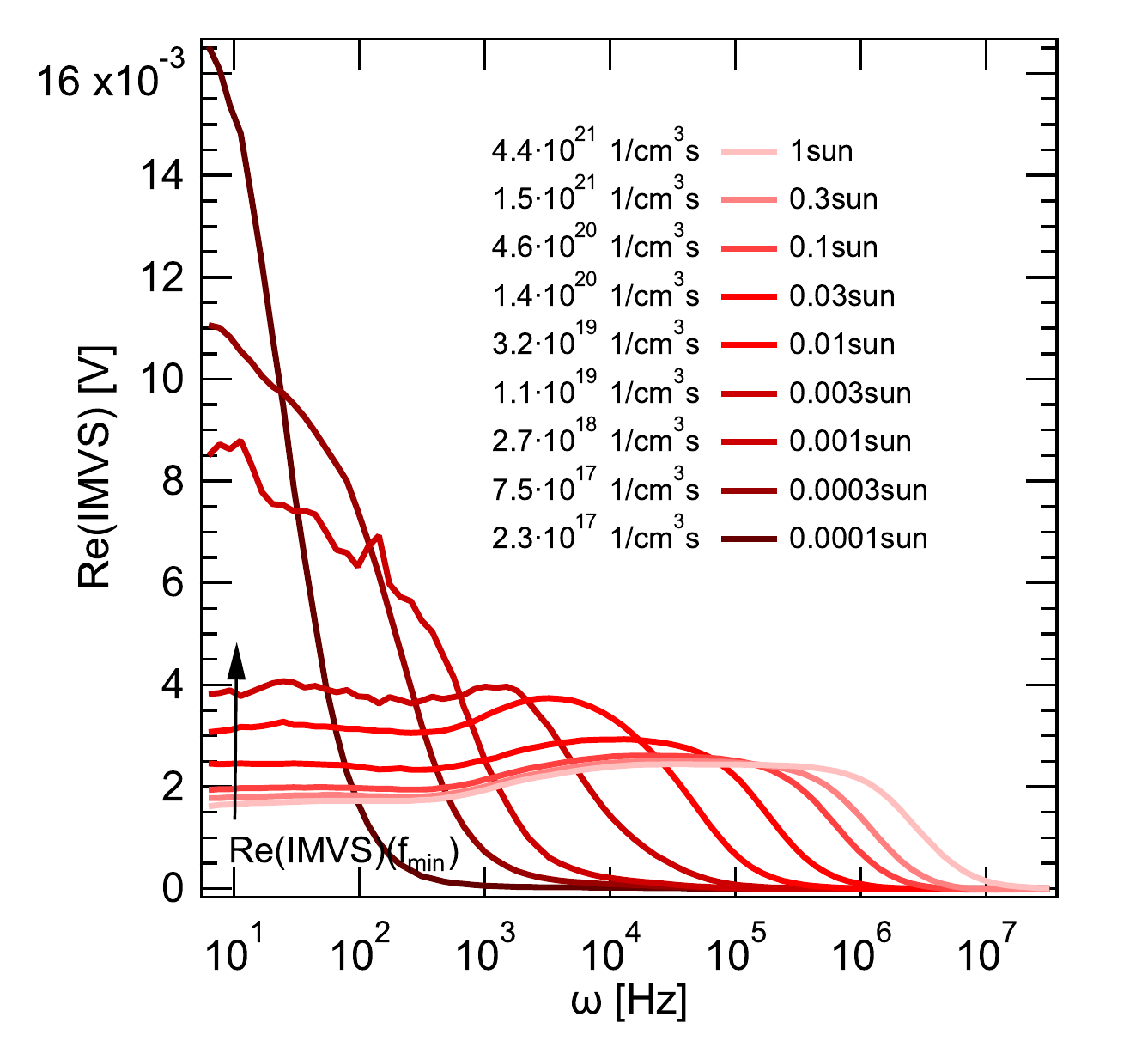}
  \caption{Real part of IMVS for different illumination intensities to determine $n_\mathrm{id}$ with Eqn.~(15).}
  \label{fig:Re_IMVS}
\end{figure}

\end{document}